\documentclass[10pt]{article}
\usepackage{geometry}                
\usepackage[parfill]{parskip}
\usepackage{amssymb}

\usepackage{amsmath}

\title{Comment on WKB Series Of All Orders}
\author{C. V. Sukumar \\{\em Wadham College,}\\{\em University of Oxford, Oxford OX1 3PN, U.K. }
}

\begin{document}
\maketitle

\begin{abstract}
The Dunham expansion for the one-dimensional two-turning-point eigenvalue problem for all orders in the WKB approximation is examined. An explicit form for all the odd order terms in the expansion which are are total derivatives is given.
\end{abstract}

\section{Introduction}
The following discussion closely follows the work of Bender {\it{et al}} \cite{BOW}. 
We consider the two-turning-point eigenvalue problem for the one-dimensional Schr{\"o}dinger equation
\begin{equation}
\Big[-\frac{d^2}{dx^2} +V(x) -E\Big] \psi(x) = 0,\quad\quad \psi(\pm\infty) =0 \label{}
\end{equation}
To construct the WKB approximation for all orders we introduce a small parameter $\epsilon$ and consider the eigenvalue problem
\begin{equation}
\epsilon^2\frac{d^2}{dx^2}\psi(x)=Q(x) \psi(x) ,\quad Q(x)= V(x) -E \label{eq:I1}
\end{equation}
The parameter $\epsilon$ helps to organize the WKB series and we can set $\epsilon=1$ at the end of the calculation. The WKB approximation to the wave function is
\begin{equation}
\psi(x)= \exp\Big[\frac{1}{\epsilon} \sum_{n=0}^{\infty}\epsilon^n S_n(x)\Big] \label{eq:I2}
\end{equation}
Substitution of this expression for $\psi$ in to the differential equation and comparison of like powers of $\epsilon$ gives
\begin{align}
[S_0^{\prime}(x)]^2 = Q(x)\ \longrightarrow\ S_0^{\prime}(x) &=-\sqrt{Q(x)} \label{eq:I3}\\
2S_0^{\prime}S_n^{\prime}+\sum_{j=1}^{n-1} S_j^{\prime}S_{n-j}^{\prime}+S_{n-1}^{\prime\prime} &= 0\ , \quad n\ge 1 \label{eq:I4}
\end{align}
where the prime symbol refers to derivatives with respect to x. The recursion relation in (\ref{eq:I4}) may be used to find $S_n^{\prime}$ from $S_j^{\prime}$ for $j < n$. The exact quantization formula for the eigenvalues valid for all orders of the WKB approximation, obtained by setting $\epsilon=1$,  has been shown to be
\begin{equation}
\frac{1}{2i}\oint_c \sum_{n=0}^{\infty} S_n^{\prime}(z) dz =K\pi,\quad K=0,1,2,... \label{eq:I5}
\end{equation}
where the integral is a complex contour integral which encircles the two turning points on the real axis. This beautiful formula was first written in this form by J.L.Dunham \cite{D}.

By direct calculation it may be shown that the first three terms of the $S_n^{\prime}$ are given by
\begin{align}
S_1^{\prime} &= -\frac{Q^{\prime}}{4Q} = -\frac{1}{4} \frac{d}{dx} [\log Q] \label{eq:I6} \\
S_2^{\prime} &= \frac{5[Q^{\prime}]^2}{32[Q]^{\frac{5}{2}}} \ -\ \frac{Q^{\prime\prime}}{8[Q]^{\frac{3}{2}}} \label{}\\
S_3^{\prime} &= \frac{-15[Q^{\prime}]^3}{64Q^4}  + \frac{9Q^{\prime} Q{\prime\prime}}{32 Q^3} - \frac{Q^{\prime\prime\prime}}{16Q^2}\quad=\ \frac{d}{dx} \Big[\frac{5[Q^{\prime}]^2}{64Q^3}  - \frac{Q^{\prime\prime}}{16 Q^2}\Big] =\ -\frac{1}{2}\frac{d}{dx}\Big[\frac{S_2^{\prime}}{S_0^{\prime}}\Big] \label{}
\end{align}

The lowest order WKB approximation arises from the use of the $n=0$ and $n=1$ terms given in (\ref{eq:I3}) and (\ref{eq:I6}) in (\ref{eq:I5}). The contribution of the $n=0$ term to the integral in (\ref{eq:I5}) may be found by the choosing a single-valued $Q(z)$ by the joining of the turning points $x_1$ and $x_2$ by a branch cut which gives 
\begin{equation}
\frac{1}{2i} \oint-\sqrt{[V(z)-E]} dz = \int_{x_1}^{x_2} \sqrt{[E-V(x)]} dx \label{}
\end{equation}
The contribution of the $n=1$ term to the integral in (\ref{eq:I5}) is 
\begin{equation}
\frac{1}{2i} \oint-\frac{Q^{\prime}(z)}{Q(z} dz = -\frac{1}{8i} [\ln Q(z)] \vert_z^z\ \Longrightarrow -\frac{1}{8i} 4\pi i= -\frac{\pi}{2} \label{}
\end{equation}
Evaluating $\ln Q(z)$ once around the contour gives $4i\pi$ because the contour encloses two simple zeros of $Q(z)$ at $x_1$ and $x_2$. Thus the first two terms in the expression given in (\ref{eq:I5}) leads to the approximate WKB quantization rule
\begin{equation}
\int_{x_1}^{x_2} \sqrt{[E-V(x)]} = \Big(K+\frac{1}{2}\Big)\pi, \quad K=0,1,2,... \label{}
\end{equation}
an approximation that increases in accuracy as $K$ increases to high values. The exact quantization formula given in (\ref{eq:I5}) is a generalization of the well known WKB quantization formula.

It has been suggested (Bender {\it{et al}}) that only the even terms in the series expansion in the Dunham formula need to be considered since $S_{2n+1}^{\prime},\ n\ge 1$, is a total derivative of a function and hence gives a vanishing contribution to the contour integral. To understand this we note that the quantization condition (\ref{eq:I5}) is a constraint on the phase of $\psi(x)$ in (\ref{eq:I2}). $S_{2n+1}^{\prime},\ n\ge 1$, is always real because it contains no fractional powers of $(V-E)$ and therefore cannot contribute to the phase of $\psi(x)$. It is $S_{2n}^{\prime}$ which becomes imaginary as x crosses into a classically allowed region and causes the wave function to become oscillatory. It is then no surprise that $S_{2n+1}^{\prime},\ n\ge 1$, drops out of the quantization condition.
Even though this reasoning by Bender {\it{et al}} is valid no direct construction showing that $S_{2n+1}^{\prime}$ is a total derivative has been given. It is my purpose to exhibit the explicit structure of $S_{2n+1}$.

\section{Odd members of the WKB series as exact derivatives}

In terms of the functions defined by $T_n\equiv S_n^{\prime}$ the recursion relation arising from the WKB series may be written in the form
\begin{align}
T_0 &= -\sqrt{Q}\ ,\quad  T_1=-\frac{T_0^{\prime}}{2T_0}\ ,\quad T_2=-\frac{T_1^{\prime}+T_1^2}{2T_0}   \label{eq:B1}\\
T_n &= -\frac{1}{2T_0} \Big[T_{n-1}^{\prime}\ +\ \sum_{m=1}^{n-1} T_m T_{n-m} \Big] \ , n\ge 3\label{eq:B2} 
\end{align}
Using the expression for $T_1$ this recursion relation may be be brought to the form 
\begin{equation}
T_n=-\frac{1}{2} \Big[\frac{d}{dx}\frac{T_{n-1}}{T_0} + \frac{1}{T_0} \sum_{m=2}^{n-2} T_m T_{n-m}\Big]\, \quad n\ge 3 \label{eq:B3}
\end{equation}
Now the even and odd members of the series can be scaled differently using 
\begin{equation}
F_j \equiv 2T_{2j+1}, \quad G_j \equiv -\frac{T_{2j}}{T_0} \label{}
\end{equation}
so that the recursion relation for the odd members of the WKB series becomes
\begin{equation}
F_n = G_n^{\prime} + \sum_{m=1}^{n-1}G_m F_{n-m} \label{eq:A1}
\end{equation}
Our aim is to prove that $F_n$ are exact differentials for all values of $n$.

The first few expressions given by
\begin{align}
F_1 &= G_1^{\prime} \label{} \\
F_2 &= G_2^{\prime} + G_1 F_1 = G_2^{\prime} + G_1 G_1^{\prime} = \frac{d}{dx} \Big[ G_2 + \frac{1}{2} G_1^2 \Big]\label{}\\
F_3 &= G_3^{\prime} + G_1 F_2 + G_2 F_1 = G_3^{\prime} + G_1 G_2^{\prime}  + G_2 G_1^{\prime} + G_1^2 G_1^{\prime} \label{}\\
&= \frac{d}{dx}\Big[ G_3 + G_1 G_2 + \frac{1}{3} G_1^3 \Big] \label{}\\
F_4 &= G_4^{\prime} + G_1 F_3 + G_2 F_2 + G_3 F1 \notag \\
    &= G_4^{\prime} + G_1 G_3^{\prime} + G_2 G_2^{\prime} + G_3 G_1^{\prime} + G_1^2 G_2^{\prime}+ G_1 G_2 G_1^{\prime} + G_2 G_1 G_1^{\prime} + G_1^3 G_1^{\prime}\label{} \\
&= \frac{d}{dx} \Big[ G_4 + G_1 G_3 + \frac{1}{2} G_2^2 + G_1^2 G_2 + \frac{1}{4} G_1^4 \Big] \label{}
\end{align}
suggest a general structure. This structure may be extracted by iteration of (\ref{eq:A1}). The first few iterations 
\begin{align}
F_n = G_n^{\prime} &+ \sum_{m=1}^{n-1}G_m \Big[G_{n-m}^{\prime} +\sum_{j=1}^{n-m-1} G_j F_{n-m-j}\Big] \notag \\
= G_n^{\prime} &+ \sum_{m=1}^{n-1}G_m G_{n-m}^{\prime} + \sum_{m=1}^{n-2} \sum_{j=1}^{n-m-1} G_m G_j \Big[G_{n-m-j}^{\prime}+ \sum_{k=1}^{n-m-j-1} G_k F_{n-m-j-k}\Big] \notag \\
=G_n^{\prime} &+ \sum_{m=1}^{n-1}G_m G_{n-m}^{\prime} + \sum_{m=1}^{n-2} \sum_{j=1}^{n-m-1} G_m G_j G_{n-m-j}^{\prime} \notag \\
&+ \sum_{m=1}^{n-3} \sum_{j=1}^{n-m-1}\sum_{k=1}^{n-m-j-1} G_m G_j  G_k F_{n-m-j-k}
\end{align}
show that the solution for $F_n$ may be given in the form
\begin{align}
F_n=G_n^{\prime}\ &+\ \sum_{m=1}^{n-1} G_m G_{n-m}^{\prime}\ +\ \sum_{m=1}^{n-2}\sum_{j=1}^{n-m-1}G_m G_j G_{n-m-j}^{\prime} \notag\\
&+\ \sum_{m=1}^{n-3}\sum_{j=1}^{n-m-1}\sum_{k=1}^{n-j-m-1}G_m G_j G_k G_{n-m-j-k}^{\prime}
\ +....+\ G_1^{n-1} G_1^{\prime} \label{}
\end{align}

It can be shown that this expression for  $F_n$ is an exact differential. This can be accomplished by noting that each of the terms in the groupings of term in the expression for $F_n$ is an exact differential because of the indexes in the summations. For example
\begin{align}
\sum_{m=1}^{n-1} G_m^{\prime} G_{n-m} &=\sum_{m=1}^{n-1} G_m G_{n-m}^{\prime} = \frac{1}{2}\frac{d}{dx}\Big[\sum_{m=1}^{n-1} G_m G_{n-m}\Big] \label{}\\
\sum_{m=1}^{n-2} \sum_{j=1}^{n-m-1} G_m G_j G_{n-j-m}^{\prime}&= \frac{1}{3} \frac{d}{dx} \Big[\sum_{m=1}^{n-2}\sum_{j=1}^{n-m-1} G_m G_j G_{n-j-m}\Big]\label{}
\end{align}
Hence
\begin{align}
F_n=\frac{d}{dx} \Big[G_n &+\frac{1}{2} \sum_{m=1}^{n-1} G_m G_{n-m} +\frac{1}{3}\sum_{m=1}^{n-2}\sum_{j=1}^{n-m-1}G_m G_j G_{n-m-j} \notag \\
&+ \frac{1}{4}\sum_{m=1}^{n-3}\sum_{j=1}^{n-m-1}\sum_{k=1}^{n-j-m-1}G_m G_j G_k G_{n-m-j-k} +....+\frac{1}{n} G_1^n \Big] \label{}
\end{align}
which explicitly demonstrates the exact differential character of $F_n$.


\begin{thebibliography}{99}

\bibitem{BOW} Bender C.M, Olaussen K., Wang P.S, Phys. Rev. {\bf{D 16}} (1977) 1710.

\bibitem{D} Dunham J.L., Phys. Rev. {\bf{41}} (1932) 713.

\end{thebibliography}
\end{document}